\documentclass[11pt]{elsart}
\usepackage{epsf}
\usepackage{float}

\begin{document}

\newcommand{\beq}{\begin{equation}}
\newcommand{\eeq}{\end{equation}}
\newcommand{\beqa}{\begin{eqnarray}}
\newcommand{\eeqa}{\end{eqnarray}}
\renewcommand{\thefootnote}{\#\arabic{footnote}}
\newcommand{\ve}{\varepsilon}
\newcommand{\krig}[1]{\stackrel{\circ}{#1}}
\newcommand{\barr}[1]{\not\mathrel #1}
\newcommand{\no}{\nonumber}
\newcommand{\dfrac}{\displaystyle{\frac}}
\renewcommand{\arraystretch}{1.1}
\hyphenation{re-nor-ma-li-zation}%
\newcommand{\vs}{\vspace{-0.25cm}}
\newcommand{\fps}{F_\pi^2}
\newcommand{\tc}{{\mathcal{T}}}
%
\begin{frontmatter} 

{\tiny \hfill FZJ-IKP(TH)-2001-06}

\title{Elastic electron--deuteron scattering \\
in chiral effective field theory}

\author[FZJ]{Markus Walzl\thanksref{MW}}, 
\author[FZJ]{Ulf-G. Mei{\ss}ner\thanksref{UGM}}
\address[FZJ]{Forschungszentrum J\"ulich, Institut f\"ur Kernphysik
  (Theorie) \\ 
D--52425 J\"ulich, Germany}
\thanks[MW]{Email: m.walzl@fz-juelich.de} 
\thanks[UGM]{Email: u.meissner@fz-juelich.de} 

\begin{abstract} 
We calculate elastic electron--deuteron scattering in a 
chiral effective field theory approach for few--nucleon systems based
on a modified Weinberg power counting. We construct the current
operators and the deuteron wave function at
next-to-leading (NLO) and next-to-next-to-leading (NNLO) order  
simultaneously within a
projection formalism. The leading order comprises
the impulse approximation of photons coupling to point-like nucleons
with an anomalous magnetic moment. At NLO, we include 
renormalizations of the single nucleon operators. To this order, no 
unknown parameters enter. At NNLO, one four--nucleon--photon operator
appears. Its strength can be determined from the deuteron magnetic moment.
We obtain not only a satisfactory description of the deuteron
structure functions and form factors measured in electron--deuteron scattering but also
find a good convergence for these observables.

\noindent
{\it PACS:} 13.40.Gp, 13.75.Cs, 12.39.Fe

\noindent
{\it Keywords:} Electron--deuteron scattering, chiral effective field theory
\end{abstract} 
\end{frontmatter}


\noindent {\bf 1.} A new era of nuclear physics calculations was started
by Weinberg~\cite{WeinI,WeinII} applying effective field theory (EFT) methods and
chiral Lagrangians to systems of two and more nucleons. Although there
are still some (minor) conceptual problems in the precise formulation of  nuclear
effective field theory, it is by now established that at very low energies,
one can perform very accurate calculations using a theory of non--relativistic
nucleons, whose interactions are given in terms of A-nucleon terms (A$=4,6,\ldots$)
with the pions integrated out, the so--called pionless theory. Going to higher
energies, the inclusion of pions becomes of prime importance and it has been
shown that Weinberg's original proposal of constructing an irreducible N--nucleon
(N$=2,3,\ldots$) potential and iterating it in a Schr\"odinger (Lippmann-Schwinger)
equation can give a precise description of nucleon--nucleon
(NN) scattering as well as static and dynamic properties of three- and four-nucleon systems, 
see e.g.~\cite{CRvK,EGMII,EIII}.
The more elegant formulation of Kaplan, Savage and Wise (KSW)~\cite{KSW}, which allows
for  power counting on the level of the scattering amplitudes, suffers in its
present formulation from the incorrect description of the tensor force, 
as reflected in the non--convergence of the triplet partial waves in $np$ scattering, 
see~\cite{MFS,RS}. For a much more detailed discussion of the concepts and 
applications of nuclear EFT, we refer to the recent reviews~\cite{birarev,border}. Of course,
many results found in nuclear effective field theory have previously been obtained in more
conventional meson-exchange approaches~\cite{DOR,CS}. These, however, cannot be
formulated in a truly systematic fashion and cannot be linked simply to the
symmetries of QCD, as it is the case of the chiral effective field theory
employed here\footnote{An illustrative example are isoscalar
  two--meson--exchange currents. While in EFT these have to be small
  simply due to power counting arguments, in conventional
  meson--exchange models such type of suppression can only be found
  after often tedious calculations, see e.g.~\cite{GM}.}.

In this letter, we will consider elastic electron--deuteron scattering based
on a Hamiltonian approach to Weinberg's formulation as developed in~\cite{EGMI}.
The central object will be the (unpolarized) scattering cross section,
which in case of the deuteron is given in terms of two structure functions,
\beq\label{XS}
\frac{d\sigma}{d\Omega} = \left(\frac{d\sigma}{d\Omega}\right)_{\rm
  Mott} \, \left[ A(q^2) + B(q^2) \tan \frac{\theta}{2} \right]~,
\eeq
with $q^2 = -Q^2 <0$ the invariant momentum transfer squared and $\theta$ is the
scattering angle in the centre-of-mass frame. Furthermore, we have separated
the QED (Mott) cross section. Alternatively, one can parameterize the
response of the deuteron to an external vector current in terms of three
form factors, $F_C$ (charge), $F_M$ (magnetic) and $F_Q$ (quadrupole).
These latter three have been the subject of a detailed study in~\cite{pc},
which employed external wave functions sewed to the leading order
interaction kernel\footnote{Earlier related work
was performed by Rho and collaborators~\cite{PMR}, who worked out the
current operators to next--to--next--to--leading
order (NNLO) and studied in particular deuteron
disintegration.}. Our approach is similar,
only that we construct the current operators and the wave functions {\em
simultaneously} from the same Hamiltonian. Furthermore, we wish to study
the interplay of nuclear and nucleon dynamics beyond the accuracy than it was done
in~\cite{pc}.  In addition, we will be able to compare our results
directly with the pioneering perturbative calculation of~\cite{KSWd},
which certainly can be considered  a cornerstone in the development of nuclear effective
field theory. Finally, this study is to be understood as the first step in
a systematic investigation of the electromagnetic properties of light nuclei.
Since much is known about the deuteron from the experimental and the theoretical side,
such an investigation is clearly needed. More precisely, the deuteron
is a very special nucleus due to its small binding energy, it thus can
be well described by iterated one--pion--exchange and some short range physics, as it
is done in conventional nuclear physics since a long time and has been
perfected to high precision. The approach presented here is similar
but also differs because corrections can be calculated in a controlled
and systematic fashion. This becomes more important if one goes to more
densely bound systems. For a more detailed discussion about the
usefulness of EFT in the description of deuteron dynamics and the
relation to conventional approaches, we refer to~\cite{pc}.


\medskip \noindent {\bf 2.} First, we briefly explain
the central ideas underlying our
calculations.  One starts from an effective chiral Lagrangian  of 
pions and nucleons, including in particular local
four--nucleon interactions which describe the short range part of 
the nuclear force, symbolically
\beq
{\mathcal L}_{\rm eff}  = {\mathcal L}_{\pi\pi} + {\mathcal L}_{\pi N} +
 {\mathcal L}_{NN}~,
\eeq
where each of the terms admits an expansion in small momenta and quark (meson)
masses. To a given order, one has to include all terms consistent with chiral
symmetry, parity, charge conjugation and so on. From the effective Lagrangian, one
derives the two--nucleon potential. This is based on a modified Weinberg 
counting, which is applied to the 
 two--nucleon potential to a certain order in small momenta and pion masses,
\beq 
V(\vec{p},\vec{p}\,') = \sum_i V^{(i)} (\vec{p},\vec{p}\,')~,
\eeq
with $\vec{p},\vec{p}\,'$ the nucleon centre-of-mass  momenta 
and the superscript
$i$ gives the (non--negative) chiral dimension. The power counting
underlying this potential is based on the considerations presented 
in~\cite{EGMI}. To leading order (LO), this
potential is the sum of one--pion exchange (OPE)
(with  point-like coupling)  and of
two four--nucleon contact interactions without derivatives. The
low--energy constants (LECs) accompanying these terms have to be
determined by a fit to some data, like e.g. the two S-wave phase shifts in the
low--energy region (for $np$). At
next--to--leading order (NLO), one has corrections to the OPE, the leading
order two--pion exchange graphs and seven dimension
two four--nucleon terms with unknown LECs (for the $np$ system).  
Finally, at NNLO, one has further corrections in the one-- and
two--pion exchange graphs including dimension two pion--nucleon
operators. The corresponding LECs can be determined from the chiral
perturbation theory (CHPT)
analysis of pion--nucleon scattering (for details, see ~\cite{EGMII}).
The existence of shallow nuclear bound states (and large scattering lengths)
forces one to perform an additional nonperturbative resummation. This is done
here by obtaining the bound and scattering states from the solution of a regularized 
Lippmann--Schwinger equation. The potential has to be understood as 
regularized, as dictated by the EFT approach employed here, i.e.
$V( {p}, {p}'\,) \to f_R ( {p} ) \, V( {p}, {p}') \, f_R ({p}' )$,
where $f_R (p)$ is a regulator function chosen in harmony with the
underlying symmetries. Within a certain range of cut--off values, the physics should be
independent of its precise form and  value. This range increases as one goes to
higher orders, as demonstrated explicitly for the $np$ case in~\cite{EGMII}.

\medskip \noindent {\bf 3.} A powerful method to simultaneously construct wave functions
and current operators was invented long time ago by Okubo and others~\cite{Okubo,FST}.
The extension to meson--exchange currents in few--nucleon systems was pioneered
by Gari and Hyuga~\cite{GH1,GH2}. As mentioned before, in~\cite{EGMI} it was shown how 
this method could be extended to chiral effective Hamiltonians based on Weinberg's 
power counting. The corresponding electromagnetic current,
denoted by $j_\mu$ throughout,
can now be obtained in two ways. One can either gauge the effective Hamiltonian or calculate
directly the effective current from the relation
\beq
\langle \psi_j | j_\mu | \psi_i \rangle  =
\langle \phi_j | j_\mu^{\rm eff} | \phi_i \rangle~,
\eeq
with $|\phi\rangle$  the projected low--energy state,
$|\psi\rangle = (1+A) ( \eta(1 + A^\dagger A)^{-1/2} \eta ) |\phi\rangle$.
Here, $\eta$ is the projector onto the Fock space with two nucleons and no 
pions and the operator $A$ can be obtained recursively from demanding that the
wave function orthogonal to $\eta |\psi\rangle$ decouples. 
The effective current takes the form  
\beq
\label{jeff}
j_\mu^{\rm eff} = \eta \, (1+ A^\dagger A)^{-1/2} (1 + A^\dagger) \,
  j_\mu \, (1+A) (1+ A^\dagger A)^{-1/2}\, \eta~.
\eeq
We mention in passing that in this approach one has to include
so--called wave function reorthonormalization diagrams, see
e.g.~\cite{EGMI} for the two--nucleon potential. Their contribution
to the effective current is completely cancelled by the two-body
recoil current (in the non-relativistic limit)~\cite{GH1}.
Before giving the explicit form of the effective current, 
it is important to formulate the power counting
in the presence of external electroweak fields, first worked out by Rho~\cite{mannque}.
Denoting by $Q$ a small external momentum or a pion mass, any QCD matrix-element
(in the presence of external fields) takes the form
\beq
{\mathcal M} = Q^\nu \, f\left(\frac{Q}{\mu},g \right)
\eeq
where $\mu$ is some regularization scale, $g$ denotes a collection
of coupling (low--energy) constants and $f$ is a function of  order one. 
The counting index is given by
\beq
\label{count}
\nu = 1 -2C +2L + \sum_i \bar{\nu}_i~, \quad \bar{\nu}_i = d_i +
\frac{n_i}{2} + e_i -2 ~,
\eeq
with $C$ the number of connected pieces in the diagram under consideration,
$L$ the number of (pion) loops, and $ \bar{\nu}_i$ is the vertex dimension. The latter
depends on the number of derivatives/pion mass insertions $(d_i)$, the number
of nucleon fields $(n_i$) and the number of external fields ($e_i)$, in our
case insertions of the electromagnetic current. Chiral symmetry
demands that the counting index is bounded from below, as verified by eq.(\ref{count}).
The terms with $C=2$ are
called one--body terms in the nuclear physics language. They subsume all interactions
of the photon with either the neutron or the proton (this is also called
the impulse approximation). For $C=1$, obviously both nucleons are involved
in the interaction, thus one talks of two--body terms (in Weinberg's language,
these terms are denoted as three--body interactions). From eq.(\ref{count}),
one quickly establishes that the lowest order terms are of the one--body
type and have $\nu =-3$ (electric photon coupling) and $\nu = -2$ (magnetic coupling).
These comprise the coupling of a photon to point--like nucleons with an anomalous
magnetic moment and define the leading order.
At NLO (and NNLO), we have further one--body terms, which are the usual leading one--loop
(third order) chiral perturbation theory corrections to the nucleon form factors,
see~\cite{BKKM,BFHM,KM}. In addition, we have the leading two--body
operators, the celebrated meson--exchange  (seagull and pionic)
currents. Using the leading order vertices as given below, these do
not contribute to elastic e-d scattering (as it is well known). Also,
based on this counting, one has no local four--nucleon--photon interactions
at NLO.  The first correction, which is not of one--body nature, 
does appear at NNLO, it is
the magnetic four--nucleon--photon interaction of~\cite{KSWd}. At that
order, one has also one--body corrections proportional to the magnetic
radii of the proton and the neutron. We should point out one subtlety
with the power counting here. In terms of the wave functions, the
current operators at NLO and NNLO are only sensitive to the NLO wave
functions, which can be simply understood from the fact that gauging
the corresponding operators in the potential does lead to even higher
orders in the current. Therefore, the results for the electric and the
quadrupole form factor at NLO and NNLO will coincide, and we denote
them as (N)NLO. Formally, one could therefore subsume what we
call NLO and NNLO here simply as NLO, similar to the two
orders making up the LO (impulse) contribution. Since this has not
be done in the existing literature, see e.g. ref.~\cite{PMR}, we
refrain from doing so too.

\medskip\noindent
We now discuss the effective interaction Hamiltonians including the photon
field, expressed in terms of the gauge potential $A_\mu$ and the field
strength tensor $F_{\mu\nu}$. From these, the effective current follows.
Gauging the free and the interaction Hamiltonians, ${\mathcal H}_0$ and
${\mathcal H}_{\rm int}$, respectively, gives (we only show the terms that involve the
photon field, all other terms to the order we are working
are given e.g. in~\cite{EGMI}), 
\beqa\label{H0}
{\mathcal H}_0^{\rm em} &=& 
- N^\dagger\left( -i \frac{\hat{e}_N}{m_N} {A}_i \, {\nabla}^i + 
 \hat{e}_N A^0  
+ \frac{e \, \kappa_s}{2m_N} 
\,  \sigma_i \, \epsilon^{ijk} \,
F_{ik} + \ldots \right) N \no \\
&& \quad +   \left( \left(\frac{\partial}{\partial t}  \pi_a \right) i \hat{e}_\pi
  A^0 \pi^a \right) - \left(\left(  \nabla_i  \pi_a \right) i \hat{e}_\pi A^i
  \pi^a \right)~, \\
{\mathcal H}_{\rm int}^{\rm em}  &=& \frac{g_A}{2 F_\pi} \, N^\dagger {\tau}_a
 \, \sigma_i (-i \hat{e}_\pi A^i) \, {\pi}^a \, N -e L_2 \, (
N^\dagger \,  \sigma_i \, \epsilon^{ijk} \, F_{ik} \, N) (N^\dagger N)~,
\label{Hint}
\eeqa
with $\hat{e}_N$ and $\hat{e}_\pi$ the nucleon and pion charge matrix,
respectively, and $m_N$ is the nucleon mass. We omit terms of higher
order in the electromagnetic coupling since they are not relevant here.
Throughout, we work in the Coulomb gauge $\vec{\nabla} \cdot \vec{A} =0$.
For the gauged free nucleon Hamiltonian, we have only given the leading order
(dimension one) electric coupling and the (dimension two)  magnetic
one, with $\kappa_s$ the isoscalar anomalous magnetic moment of the nucleon,
$(\kappa_p +\kappa_n)/2 = \kappa_s$. These two interactions lead to
the terms with $\nu =-3$ and $-2$ as discussed before. The dimension
three terms contributing to the one--body currents at $\nu =-1$ can be
found in~\cite{BFHM}.  The pion--photon--nucleon interaction in
eq.(\ref{Hint}) is nothing but the celebrated Kroll-Ruderman term,
with $g_A$ the axial vector coupling measured in neutron
$\beta$--decay and $F_\pi$ the weak pion decay constant.
As noted before, employing these Hamiltonians, the leading exchange
currents are proportional to $\vec{\tau}_1 \times \vec{\tau}_2$ 
(where $\vec{\tau}_i$ is
the isospin operator of nucleon $i$) and thus vanish for isoscalar transitions
as it is the case for e-d scattering. The leading isoscalar exchange
current appears only at N$^3$LO, because the only possible NNLO
term would consist of a dimension two correction to the Kroll-Ruderman
vertex. This operator has the structure $\sigma_i \epsilon^i \,
v_\mu q^\mu \, (\tau^a + \delta^{a3})$ (see app.~A of \cite{bkmrev}) and
thus generates only an isovector operator.  At N$^3$LO, we have
contributions from two--pion--exchange diagrams with one
$\gamma \pi\pi \bar NN$--vertex (football and triangle
graphs) together with dimension three corrections to the
pion-nucleon-photon vertex. The last term in eq.(\ref{Hint}), first
considered in~\cite{KSWd}, is a
magnetic photon four--nucleon contact  interaction, its strength can
not be determined from np scattering but can be obtained from a fit to
the deuteron magnetic moment. We note that there is also a NNLO
single nucleon correction due to the magnetic radius.
We stress that such power counting arguments have already been given
in~\cite{pc}.

\medskip
\noindent {\bf 4.} We now turn to the response of the deuteron to an
external electromagnetic field. Consider a deuteron state with
four--momentum ${p}_\mu$ and polarization vector $\vec{\epsilon}\,^\mu$,
subject to the condition $p_\mu \vec{\epsilon}\,^\mu = 0$. In the
deuteron rest frame, one conventionally selects $\epsilon^\mu_i =
\delta_i^\mu$ and the corresponding deuteron states are denoted by
$|\vec{p} , i\rangle$. In terms of these and to leading order in the 
non-relativistic expansion, the matrix element of the electromagnetic
current is given in terms of the charge, magnetic and quadrupole
form factors,
\beqa\label{ME}
\langle \vec{p}\,' , i \, | \, j_0 \, | \, \vec{p} , j \rangle &=&
e \, \left[ F_C(Q^2) \, \delta_{ij} + \frac{1}{2m_d^2} F_Q (Q^2)
\left( \vec{q}_i \vec{q}_j - \frac{1}{3} \vec{q}\,^2 \delta_{ij}
\right)\right]~, \nonumber \\
\langle \vec{p}\,' , i \, | \, j_k \, | \, \vec{p} , j \rangle &=&
\frac{e}{2m_d} \, \left[ F_C(Q^2) \, \delta_{ij} (\vec{p} +
\vec{p}\,')_k + F_M(Q^2) (\delta_{jk}\vec{q}_i -\delta_{ik}\vec{q}_j)
\right. \no \\
&& \qquad\quad + \left. \frac{1}{2m_d^2} F_Q (Q^2)
\left( \vec{q}_i \vec{q}_j - \frac{1}{3} \vec{q}\,^2 \delta_{ij}\right)
(\vec{p} + \vec{p}\,')_k
\right]~,
\eeqa  
with $\vec{q} = \vec{p}\,' - \vec{p}$, $Q = |\vec{q}\,|$ and $m_d$ is
the deuteron mass. These
form factors are dimensionless and  normalized to the charge, the
magnetic moment, $\mu_d$, and the quadrupole moment, $Q_d$, of the deuteron, i.e.
$F_C (0) =1$, $(e/2m_d)F_M(0) = \mu_d$ and $(1/m_d^2)F_Q (0) = Q_d$,
with $\mu_d = 0.85741\, (e/2m_N)$ and $Q_d = 0.2859\,$fm$^2$.
The structure functions $A(Q^2)$ and $B(Q^2)$ defined in eq.(\ref{XS})
are given in terms of these form factors via
\beqa\label{ABF}
A &=& F_C^2 + \frac{2}{3} \eta F_M^2 + \frac{8}{9} \eta^2 F_Q^2~,
\nonumber \\ 
B &=& \frac{4}{3} \eta (1+\eta ) F_M^2~.
\eeqa
Here, $\eta =  Q^2/4m_d^2$. To disentangle these three form factors, one
measurement involving polarization is necessary. The analyzing power
$t_{20}$ has become the observable of choice to do that. Since it
depends on $Q^2$ and the scattering angle $\theta$, one also uses the quantity
$\tilde{t}_{20}$, 
\beq
\tilde{t}_{20} = -\frac{\frac{8}{3}\eta F_C (Q^2) F_Q (Q^2) +
  \frac{8}{9} \eta^2 F_Q^2 (Q^2)}{\sqrt{2} \left[F_C^2 (Q^2) +
 \frac{8}{9} \eta^2 F_Q^2 (Q^2)\right] }~,
\eeq
which only depends on the momentum transfer squared and is independent
of the magnetic form factor. Note, however, that $\tilde{t}_{20}$ is
not directly measurable (for a detailed discussion, we refer to the
recent review~\cite{drev}).
 
\medskip

\noindent {\bf 5.} We first consider the structure functions $A$ and
$B$ which can be obtained directly from elastic e-d scattering. Since
from the study of the nucleon form factors in heavy baryon chiral
perturbation theory it is known that the third order result for the
isoscalar form factor starts to deviate from the data at $Q^2\simeq
0.2\,$GeV$^2$, we will limit our range of $Q$ from 0 to 400$\ldots$500~MeV.
We work out the matrix elements from eqs.(\ref{ME}) and construct the
structure functions by use of eqs.(\ref{ABF}) without any
approximation. This is different from~\cite{KSWd}, where the
kinematical factors like e.g. $\eta$ where also included in the power
counting and thus in their case there can be no contribution from
$F_Q (Q^2)$ to $A(Q^2)$ at NLO. Throughout, we work with an
exponential regulator $f_R (p) = \exp (-p^4 /\Lambda^4)$ with a cut--off
$\Lambda = 600\,$MeV (similar to what was done in the simultaneous fit to
$np$ and $pp$ scattering phases, see~\cite{WME}). None of the results
presented here depends notably on the cut--off within its allowed
bounds. For completeness, we give the corresponding LECs. At LO,
we have $\tilde{C}_{3S1} = 0.0427\cdot 10^4\,$GeV$^{-2}$ and at NLO,
we get $\tilde{C}_{3S1} = -0.0368\cdot 10^4\,$GeV$^{-2}$,
${C}_{3S1} = 0.186\cdot 10^4\,$GeV$^{-4}$ and
${C}_{3S1-3D1} = -0.190\cdot 10^4\,$GeV$^{-4}$. These numbers are in
good agreement with what has been found before using such type
of regulator, see e.g. ref.~\cite{EEthesis}.

\medskip

\begin{figure}[htb]
\centerline{
\epsfxsize=9cm
\epsffile{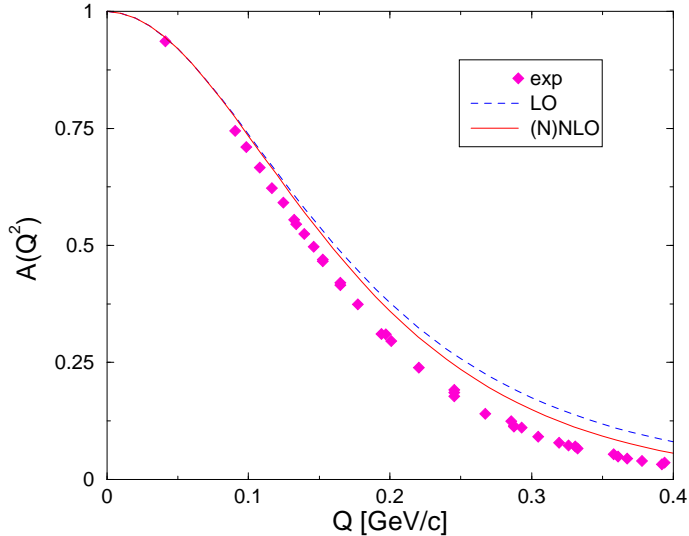}
}
\vspace{-0.1cm}
\caption{The structure function A($Q^2$). The dashed and the solid line
  give the LO, the (N)NLO result, respectively, and the data are from
  \protect\cite{simon}, \protect\cite{platchkov}. As explained in the
  text, the  NLO and NNLO results for $A(Q^2)$ can not be
  distinguished on the scale of the figure.\label{fA}}
\vspace{0.3cm}
\end{figure}

\noindent Let us  discuss $A(Q^2)$ as given in fig.~\ref{fA}.
Note that we plot the structure function versus $Q \equiv \sqrt{Q^2}$
(not versus $Q^2$) to facilitate the comparison with ref.~\cite{KSWd}.
The leading order curve already gives a fair description of the data,
the NLO and NNLO result is visibly improved but still above the data.
Note that the difference between NLO and NNLO stems from a Foldy--type contribution
proportional to the magnetic radius, which is very small, and an
induced contribution from the magnetic photon--four--nucleon
interaction. Both of these are small and further suppressed by a
factor of $\eta$, so that the
difference between the two curves can not be seen on the scale of fig.~\ref{fA}.
Since there is no exchange current contribution, the  improvement 
going from LO to NLO is due
to the single nucleon radius terms which appear at this order. 
This can be seen by considering the LO wave function together with
the NLO current operators. At yet higher orders, however, there should be some
further improvement due to the wave functions as shown in~\cite{pc}.
One should consider such a 
sensitivity to single nucleon observables not as an unwanted
complication but rather conclude that indeed isoscalar
quantities (alas neutron properties) can be inferred from nuclear
targets with some precision. We consider this interplay of chiral
nucleon dynamics and nuclear EFT as one of the important ingredients 
in the whole approach. Note further that our (N)NLO result is similar
to the one of~\cite{KSWd}. It should also be noted
that for the range of momentum transfer considered here, the NLO
corrections are sizeably smaller than in the KSW calculation, pointing towards
a better convergence.

\begin{figure}[htb]
\centerline{
\epsfxsize=9cm
\epsffile{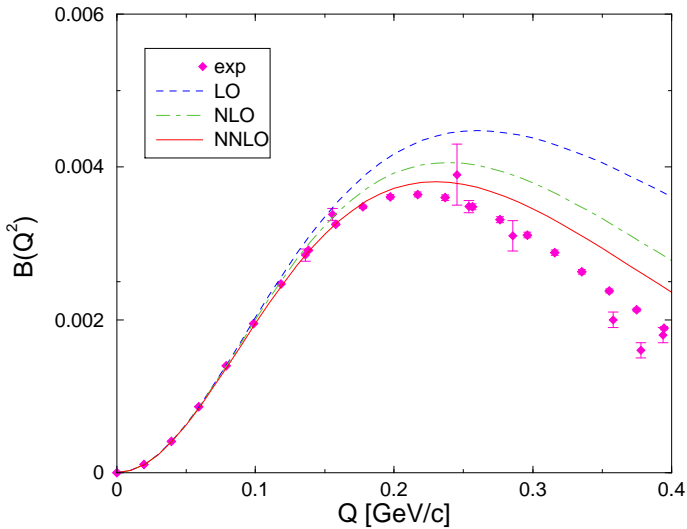}
}
\vspace{-0.01cm}
\caption{The structure function B($Q^2$). The dashed, the dot-dashed and the solid line
  give the LO, the NLO and the NNLO result, respectively. The new data are from
  the compilation of Sick~\cite{Ingo} (filled diamonds) and the
  older data (open circles) are from 
  \protect\cite{gdl}, \protect\cite{bdf}, \protect\cite{ggi}.\label{fB}}
\end{figure}

\noindent
Next, we turn our attention to $B(Q^2)$, which is
essentially the response to the magnetic photon, see fig.~\ref{fB}.
Again, the LO result
is visibly better than the one obtained in~\cite{KSWd}, whereas the
NLO curves are comparable. However, in contrast to the work
of~\cite{KSWd}, our prediction for $B(Q^2)$ is parameter--free,
whereas the one in the KSW scheme has one free parameter from the 
four--nucleon--photon interaction (due to the different counting). 
Also, the  NLO correction is fairly small for the momentum range considered here,
indicating  convergence. At NNLO, we have to pin down the LEC $L_2$ of the
magnetic photon four--nucleon interaction, which is done by fitting the
magnetic moment of the deuteron. The resulting curve for $B(Q^2)$ is
further improved,  see fig.~\ref{fB}.

\medskip

\noindent Next, we discuss the
charge, magnetic and quadrupole form factors. The corresponding 
normalizations (static properties) are collected in table~\ref{t1}
for LO and NLO. The only difference at NNLO is the magnetic moment,
which is exactly reproduced (since it is used as input to fix $L_2$).
The corresponding value for the
the LEC $L_2$ is $L_2 = 2.68\cdot 10^{-2}\,$GeV$^{-2}$, which is
rather small. This can be traced back to the fact that the NLO
result for $\mu_d$ is already within 1\% of the empirical value. 
Note that in contrast to what was done in ref.\cite{EGMII}, we have
fine tuned the LECs in the deuteron channel to reproduce the binding energy
to four digits. The only difference at NNLO to NLO is the magnetic
moment, therefore the NNLO numbers are not given. Notice that with the
fine tuned binding energy, the prediction for the quadrupole moment
is visibly improved as compared to ref.~\cite{EGMII} and closer to the
data than in all modern high--precision potentials.

\begin{figure}[htb]
\vspace{0.2cm}
\centerline{
\epsfysize=7cm
\epsffile{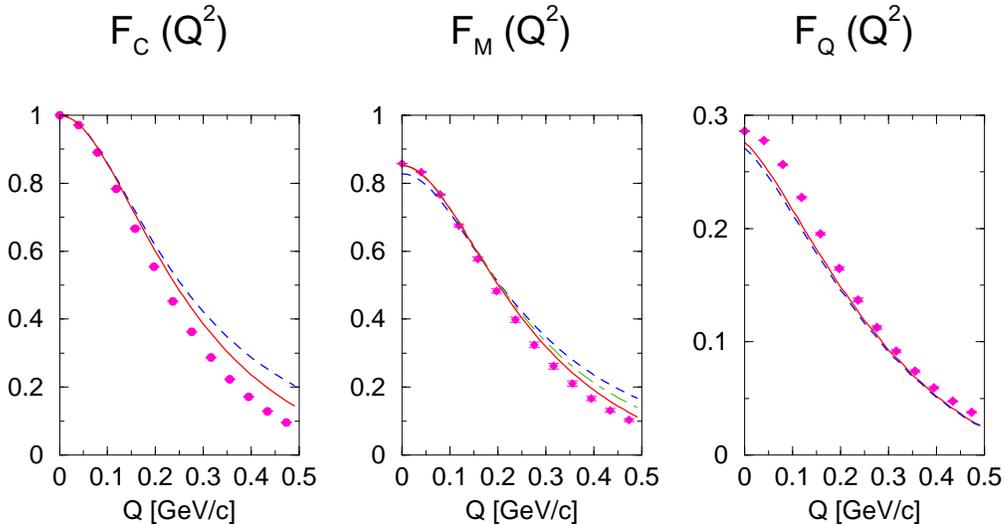}
}
\vspace{-0.01cm}
\caption{Electric, $F_C (Q^2$),  magnetic, $F_M (Q^2)$,
  and quadrupole, $F_Q (Q^2)$, form factors of the
  deuteron, in order. The dashed, the dot-dashed and the solid lines
  give the LO, the NLO (only shown for $F_M$) and the NNLO results, 
  in order, and the data are from the analysis of Sick~\cite{Ingo}.
  As explained in the text, NLO and NNLO results coincide for the
  electric and the quadrupole form factor.
  \label{fF}}
\vspace{0.3cm}
\end{figure}

\begin{table}[h]
\begin{center}
\caption{Static properties at LO and NLO. Here, $E_d$, $\mu_d$ and
  $Q_d$ denote the deuteron binding energy, its magnetic and
  its quadrupole moment. }
\smallskip
\begin{tabular}{|l|c|c|c|}
\hline
                  &  LO     &  NLO    & Exp.     \\ \hline 
$|E_d|$ [MeV]     & 2.224   & 2.224   & 2.22456612(12) \\
$\mu_d$ [$\mu_N$] & 0.828   & 0.852   & 0.8574382284(94)  \\
$Q_d$ [fm$^2$]    & 0.265   & 0.276   & 0.2859(3)  \\
\hline
\end{tabular}
\label{t1}
\end{center}
\end{table}
\noindent
In all cases, the NLO/NNLO corrections are small and one finds a 
decent description of the three form factors, see fig.~\ref{fF}. 
The largest discrepancies at higher momentum transfers
 are found in the charge form factor,
as reflected in the results for $A(Q^2)$ shown above.
Similar results have been obtained in~\cite{pc}, however, 
a direct comparison with data was not given in that paper.

\medskip

\begin{figure}[htb]
\centerline{
\epsfysize=6.5cm
\epsffile{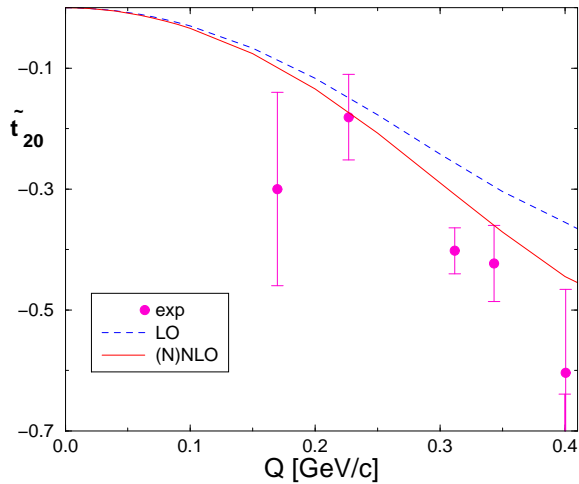}
}
\vspace{-0.01cm}
\caption{Analyzing power $\tilde{t}_{20}$ at LO (dashed) and (N)NLO
  (solid line). The data are from the recent compilation~\cite{JT20}.
  \label{fT}}
\vspace{0.3cm}
\end{figure}

\noindent
From the electric and the quadrupole form factor,
we can construct $\tilde{t}_{20}$, as shown in fig.~\ref{fT}.
Again, the deviations of our LO and (N)NLO predictions for $F_C (Q^2)$
from the data are responsible for the too small magnitude of
$\tilde{t}_{20}$ at $Q \simeq 0.4\,$GeV.

\vfill\eject

\medskip\noindent {\bf 6.} We have analyzed electron--deuteron
scattering in the framework of a chiral effective field theory
for few--nucleon systems at next--to--leading 
and next--to--next--to--leading order. 
At NLO, no meson--exchange currents or four--nucleon--photon operators
contribute. At NNLO, only one magnetic photon--four--nucleon operator
appears, the corresponding coupling constant can be determined from
a fit to the deuteron magnetic moment. As stressed, the NLO and NNLO
predictions for the electric and the quadrupole form factors coincide
at NLO and NNLO because one is not yet sensitive to the NNLO wave
function corrections. In particular, we have 
discussed the interplay between the single nucleon dynamics as 
encoded in the nucleon form factors and the nuclear dynamics. As
already stressed in~\cite{pc}, the accuracy of the description of
the nucleon form factors limits the applicability of the effective field
theory approach to the deuteron structure to momentum transfer of
about $Q \simeq 0.4$~GeV. Thus an improved description
of these single nucleon observables has to be obtained to extend these
considerations consistently to higher photon virtualities. For a step in
this directions, see~\cite{KM}. We stress again that we consider it
important to consistently describe the single as well as the
few--nucleon sector. The results presented here
extend the ones of~\cite{pc} to the first non--trivial order for the
single nucleon form factors, i.e. are sensitive to the structure of the
nucleon. Obviously, as next steps one has to
consider N$^3$LO corrections to the process discussed here
(since only at that order sufficiently many additional contributions
appear, like e.g. meson--exchange currents) 
as well as electron scattering of three- and four--body systems, using the
wave functions of~\cite{EIII}.

\vfill\eject

\noindent {\bf Acknowledgements}

\noindent
We are grateful to Manfred Gari for a useful discussion, Ingo Sick
for providing us with the e-d scattering data and his analysis of
these and Rocco Schiavilla for supplying his data base.
We thank Daniel Phillips for some clarifying remarks on the work 
presented in~\cite{pc} and Bastian Kubis for comments and 
a careful reading of the manuscript.

\end{document}